**Interacting multi-channel topological boundary modes in a quantum Hall valley system**

Mallika T. Randeria[1*], Kartiek Agarwal[2*], Benjamin E. Feldman[1†], Hao Ding[1], Huiwen Ji[3], R. J. Cava[3], S. L. Sondhi[1], Siddharth A. Parameswaran[4], Ali Yazdani[1‡]

[1]*Joseph Henry Laboratories & Department of Physics, Princeton University, Princeton, NJ 08544, USA*
[2]*Department of Electrical Engineering, Princeton University, Princeton NJ 08544, USA*
[3]*Department of Chemistry, Princeton University, Princeton, NJ 08544, USA*
[4]*The Rudolf Peierls Centre for Theoretical Physics, University of Oxford, Oxford OX1 3NP, UK*
[†]*Present address: Department of Physics and Geballe Laboratory for Advanced Materials, Stanford University, Stanford, CA 94305, USA*

[‡] email: yazdani@princeton.edu
* these authors contributed equally

**Symmetry and topology play key roles in the identification of phases of matter and their properties. Both concepts are central to understanding quantum Hall ferromagnets (QHFMs), two-dimensional electronic phases with spontaneously broken spin or pseudospin symmetry whose wavefunctions also have topological properties[1,2]. Domain walls between distinct broken symmetry QHFM phases are predicted to host gapless one-dimensional (1D) modes that emerge due to a topological change of the underlying electronic wavefunctions at such interfaces. Although a variety of QHFMs have been identified in different materials[3-8], probing interacting electronic modes at these domain walls has not yet been accomplished. Here we use a scanning tunneling microscope (STM) to directly visualize the spontaneous formation of boundary modes, within a sign-changing topological gap, at domain walls between different valley-polarized quantum Hall phases on the surface of bismuth. By changing the valley occupation and the corresponding number of modes at the domain wall, we can realize different regimes where the valley-polarized channels are either metallic or develop a spectroscopic gap. This behavior is a**

**consequence of Coulomb interactions constrained by the symmetry-breaking valley flavor, which determines whether electrons in the topological modes can backscatter, making these channels a unique class of interacting Luttinger liquids.**

The broken symmetry of QHFMs is often a purely internal electronic degree of freedom such as electron spin, valley flavor, or orbital index. However, anisotropic multi-valley electronic systems present a special case where the QHFM order parameter couples to the spatial degrees of freedom[9,10]. This gives rise to discrete nematic order, which breaks the discrete rotational symmetry of the underlying crystal lattice and hence is especially sensitive to spatial inhomogeneities, e.g. from disorder or strain. When present, these break uniform nematic quantum Hall phases into domains whose boundaries host one or more sets of valley-polarized 1D modes (Fig. 1a,b). These edge modes can be mapped to Luttinger liquids[11]; they are analogous to those studied in purely 1D systems[12-15], but with the novelty that valley flavor can dictate their properties. To date, QHFM has been observed via bulk measurements[3-8] and, very recently, with STM[16,17]. Transport studies, which measure macroscopically averaged electrical resistance, are poorly suited to address the microscopic properties of domain walls between phases. Our experimental approach is to leverage the accessibility of quantum Hall states formed on the surface of bismuth (Bi) and use STM measurements to directly visualize domain walls between different valley-polarized phases. We use the tunability of valley occupation in this system in combination with high resolution STM spectroscopy to probe the properties of valley-polarized topological boundary modes. The presence of such boundary modes are guaranteed by the Callan-Harvey anomaly cancellation mechanism[18], which allows the strongly interacting setting of our experiment to be treated with theoretical generality. Our results, combined with a recent theoretical analysis of nematic quantum Hall domain walls[19], show that these interacting systems harbor a new class of symmetry-protected Luttinger liquids.

The multi-valley system we study here is associated with the six quasi-elliptical hole valleys of the Bi(111) surface Brillouin zone (in momentum space)[20] (Fig. 1c,d). The six-fold degeneracy can be lifted in the quantum Hall regime either partially by strain, or fully by electron-electron interactions[16,17]. When interaction effects are dominant, this system can form domain walls between distinct valley-polarized states, with different number and valley flavor of the boundary modes depending on the filling factor (Fig. 1a,b). Here we explore two QHFM domain walls that emerge when either one-out-of-four or two-out-of-four degenerate valleys are occupied. We label these two cases by their effective filling factors, $\tilde{\nu}=1$ and $\tilde{\nu}=2$ respectively, where $\tilde{\nu}$ is defined as the number of occupied states within the relevant subset of valleys participating in the domain wall formation. For $\tilde{\nu}=1$, the domain wall hosts a single pair of counter-propagating modes arising from different valleys (e.g. valleys A and B; Fig. 1b). At $\tilde{\nu}=2$, there are two pairs of counter-propagating modes, and the valley flavors of each of the two co-propagating channels are also distinct (i.e. states from valleys A and $\bar{A}$ moving in one direction and states from valleys B and $\bar{B}$ moving in the other direction along the domain wall; Fig. 1a). Such domains with different orientations of the broken valley symmetry are also topologically distinct – intuitively, they have different 'valley Chern number', defined as $N_v = \nu_A + \nu_{\bar{A}} - \nu_B - \nu_{\bar{B}}$. Consequently, these boundary modes may be identified with the gapless conducting states that occur as a bulk topological invariant of the quantum Hall state changes across the domain wall, depicted schematically in Fig. 1e,f.

We use a combination of energy- and spatially-resolved spectroscopy with a high-field dilution refrigerator STM[21] to study the QHFM domains and their 1D edge modes on the Bi(111) surface. Measurements of differential tunneling conductance *dI/dV* with high energy resolution in Fig. 2a,b show examples of different broken symmetry states of the six hole valleys, where the occupation of a Landau level can be tuned by changing the magnetic field (see Methods). In the vicinity of the domain wall we describe here, strain lifts the degeneracy for two of the six valleys. Electron-electron interactions further split in energy the remaining four-fold multiplet around the Fermi level (*E*=0), either into a pair of doubly degenerate levels (corresponding to $\tilde{v}$=2; Fig. 2a), or, at a different value of the magnetic field, into one singly and one triply degenerate state ($\tilde{v}$=1; Fig. 2b). The interaction-induced exchange gap between occupied and unoccupied valleys at the Fermi level ($\Delta_{exch} \approx 650$ μeV) is a signature of spontaneous symmetry breaking. We identify the valley ordering of these quantum Hall phases and their emergent nematicity by imaging the corresponding Landau level wavefunctions near individual impurities, as we have demonstrated previously[16,17]. Here we extend this capability to visualize the interface between different nematic phases, identify the valleys that participate in the formation of the domains, and investigate the resultant boundary modes.

We locate nematic domain walls by STM spectroscopic mapping of spatial variations in *dI/dV* at an energy near the Fermi level, such as those in Fig. 2c,e that are measured in large pristine areas of the Bi(111) surface (Fig. 2d). In maps measured at the energies corresponding to the exchange-split Landau levels, the domain walls appear as a stripe of low tunneling conductance between two regions with inequivalent broken rotational symmetry, as shown in Fig. 2c,e. On either side of the domain walls, the spectroscopic maps show elliptical features corresponding to Landau orbits pinned to a very low concentration of individual atomic defect sites (see Methods). The orientations of these wavefunctions reflect the direction of the anisotropy of the hole pockets (Fig. 1d) from which they originate, indicating the specific valleys associated with each domain. Considering the orientations of the wavefunctions on either side of the domain wall in Fig. 2c together with the corresponding spectra in Fig. 2a, we conclude that there are pairs of valleys occupied in each domain, i.e. they have an effective filling factor $\tilde{\nu}=2$. Measurements taken in the same area of the sample as Fig. 2c,d but at a different magnetic field, show the $\tilde{\nu}=1$ domain wall where the occupation of a single valley switches across the boundary (schematic Fig. 1b,f). In the map measured at the higher energy of the exchange split peaks (Fig. 2e), Landau orbits of two different orientations are visible on either side of the domain boundary, but elliptical features associated with one of the orientations on either side are more pronounced than for the other, in agreement with expectations for a triply degenerate state (spectra in Fig. 2b). Additional conductance maps illustrate the energetic behavior of these nematic domains and show a switch of LL wavefunction directionality when probing the occupied vs. unoccupied valleys (see Methods). Finally, tuning the LLs away from the Fermi level or increasing temperature, conditions where exchange effects are absent or suppressed, results in the disappearance of the domains and domain walls (see Methods). From these measurements, we conclude that the nematic domains in our sample, with typical size of a few microns, are spontaneously formed due to electron-electron interactions, although local strain likely provides a small symmetry breaking perturbation that determines which valleys are occupied in each domain (see Methods).

A hallmark of topological boundary modes is that they energetically occur in the gap of the corresponding bulk states and spatially reside where a topological invariant linked to this gap changes sign at an interface between two topologically distinct phases[22]. Spatially resolved STM spectroscopic line-cuts measured across the domain walls shown in Fig. 3a,b demonstrate how the exchange gaps defining our local valley-polarized domains close and reopen. This profile of the gap (Fig. 3a,b), as well as the measurements described above of the underlying valley state wavefunctions on either side, together demonstrate the effective sign change of the topological invariant at the domain walls in our system. Spectroscopic maps obtained at the Fermi energy reveal the spatial structure of electronic states within the gaps and exhibit high tunneling conductance along the domain walls, establishing the presence of low energy boundary modes (Fig. 3c,d). Unlike the chiral edge modes along the perimeter of a quantum Hall system, we expect these topological 1D states that emerge at the boundary between valley-polarized domains to be counter-propagating. Although many topological electronic phases have been identified to date[22], no previous experimental work has demonstrated a direct link between closing an interaction-induced energy gap and a spatial "twist" in a bulk topological invariant. In contrast to domain-wall modes in bilayer graphene[23-25], which can be understood from a single-particle perspective, here the physics is entirely interaction-driven so that the domain walls host strongly-interacting Luttinger liquids[19].

Theoretically, we can understand key aspects of our experimentally observed topological edge modes between nematic quantum Hall states[19]. The highly anisotropic shape of the Landau orbits is due to the large effective mass anisotropy ($m_{//}/m_\perp \approx 25$) of the bismuth valleys[16,20] which favors an abrupt change in the valley occupation across the nematic domain wall[26]. The extent of the wavefunctions perpendicular to the domain wall determines the spatial width of the boundary modes (approximately 100nm in Fig. 3c,d). The spectra also reveal an asymmetry in the amplitude of the LL peaks near the domain wall (Fig. 3g), which is a signature of a dipole moment that arises from the different spatial extent of the Landau orbits from the two differently oriented valleys projected onto the 1D boundary[26]. Detailed numerical Hartree-Fock calculations not only capture the closing and opening of the exchange gap and the experimental width of the domain wall (Fig. 3e,f), but also account for the asymmetry in the local density of states near the domain wall due to the presence of a dipole moment (Fig 3g; see Supplementary Information).

Further investigation of the electronic behavior at the domain walls using STM spectroscopy reveals that these 1D channels can be metallic or insulating depending on the nature of valley states from which they emerge. Although we observe the presence of low energy modes along the boundary between topologically distinct nematic phases (Fig. 3a,c), individual spectra taken at the domain wall for $\tilde{v}=2$ exhibit a local charge gap of $\Delta_{\text{charge}}$ ranging from 325 µeV to 425 µeV (Fig. 4a, Methods). This $\tilde{v}=2$ domain wall is expected to contain two pairs of counter propagating 1D channels, with states from valleys A, $\bar{\text{A}}$ moving in one direction and valleys B, $\bar{\text{B}}$ moving in the opposing direction along the boundary (Fig. 1a). In contrast, at the $\tilde{v}=1$ domain wall, we find a single peak in the density of states (Fig. 4b), with no resolvable gap when the boundary is expected to host singly-degenerate counter-propagating modes (Fig. 1b). These strikingly different spectra are observed for $\tilde{v}=1$ and $\tilde{v}=2$ domain walls formed at the same location on the sample, subject to the same background of impurities. If the $\tilde{v}=2$ charge gap were induced by disorder, then we expect the same effect to also visibly gap the $\tilde{v}=1$ modes, but the latter remains gapless (see Methods for measurements at multiple locations along domain wall). We therefore attribute the formation of an insulating state not to localization from disorder, but to Coulomb interactions between 1D modes, which provides a natural explanation of why the presence or absence of an interaction gap depends on the filling factor, as we now describe.

The experimental observation of a filling factor-dependent charge gap may be understood by considering all possible four fermion interactions between the different valley-polarized modes at the domain wall. We account for both the chirality, $r$, which denotes the direction of propagation, and the flavor, $\sigma$, which distinguishes the valley origin of co-propagating states of the edge modes at different filling factors. We describe the interaction processes as $i_1 i_2 \rightarrow f_1 f_2$, where two incoming states $i_1$ and $i_2$ interact to produce the outgoing states $f_1$ and $f_2$. For $\tilde{\nu}=1$, the two relevant channels are the counter-propagating A and B edge modes which can be labeled as ($r = +1$, σ = +1) and ($r = -1$, σ = +1), respectively. Interactions of the form AB → BA (Fig. 4e) do not change the net chirality of the system, and therefore the domain wall remains gapless. Backscattering processes of the form AA→ BB are exponentially suppressed due to the large momentum transfer between these valleys in the 2D Brillouin zone (Fig. 4f). In comparison, $\tilde{\nu}=2$ has two modes propagating in each direction with opposite flavors, i.e. A and $\bar{A}$ have the same chirality $r = +1$ but different flavor indices, σ = +1 and σ = -1, respectively, and similarly the B and $\bar{B}$ modes are labeled by ($r = -1$, σ = +1) and ($r = -1$, σ = -1). In this case, the interaction A$\bar{A}$ → B$\bar{B}$ changes the chirality of the system, and does so without any net momentum transfer in the 2D Brillouin zone (Fig. 4c), thereby opening up a gap. Heuristically, such interaction-induced backscattering can be understood by treating the valley flavor analogously to spin. Our experimentally-observed dichotomy between edge modes at $\tilde{\nu}=1$ and $\tilde{\nu}=2$ is similar to that seen in spinless versus spinful Luttinger liquids, where an energy gap is only expected in the latter case[11](see Supplementary Information). In addition to the charge gap at $\tilde{\nu}=2$, a gapless neutral valley mode is also expected at the domain wall[19], and the detection of such valley-charge separation in a Luttinger liquid would be an exciting avenue to pursue with other measurement techniques.

The valley-polarized QHFM states examined here can be realized in a wide range of materials and provide the opportunity to examine not only different types of Luttinger liquids but also to explore novel ways of connecting them. Multi-valley systems such as graphene[7,27], transition metal dichalcogenides[28] or topological crystalline insulator surface states[29,30] can be used as material platforms to explore domain boundaries between a multitude of QHFM phases. These 2D systems naturally lend themselves to STM studies similar to those performed here that can both visualize electronic domain walls and probe the properties of their associated boundary modes. Moreover, the Bi(111) hole states provide an intriguing opportunity for domain structures involving all three nematic orientations. Theoretically, the possibility of such Luttinger liquid Y-junctions has been shown to have a variety of possible electronic behavior[31], whose interplay with the underlying quantum Hall and valley physics can potentially drive a rich phase structure. Finally, the application of in-situ strain to control the location of domain walls in nematic QHFMs, opens up the possibility of imaging domain wall dynamics and makes these boundary modes more accessible to other measurements, including transport studies.

**Acknowledgements**

The experiments in this project were primarily supported by the Gordon and Betty Moore Foundation as part of the EPiQS initiative (GBMF4530) (BEF, HD, AY) and DOE-BES grant DE-FG02-07ER46419 (MTR, AY). Other financial support for the experimental effort has come from NSF-DMR-1608848 (AY), NSF-MRSEC programs through the Princeton Center for Complex Materials DMR-142054 (AY, RJC), and by a Dicke fellowship (BEF). AY acknowledges the hospitality of the Aspen Center for Physics supported under NSF grant PHY-1607611. KA acknowledges support from DOE-BES Grant No. DE-SC0002140 and the U.K. foundation. SAP acknowledges support from NSF DMR-1455366 during the early stages of this project. We acknowledge useful discussions with S. Kivelson and E. Fradkin.

**Author Contributions**

M.T.R., B.E.F., H.D., and A.Y. designed and conducted the STM measurements and their analysis. K.A., S.L.S, and S.A.P performed the theoretical calculations. H.J. and R.J.C. synthesized the samples. All authors contributed to the writing of the manuscript.

**Author Information**

"Reprints and permissions information is available at www.nature.com/reprints.

The authors declare no competing financial interests.

Correspondence and requests for materials should be addressed to yazdani@princeton.edu.

**Figure captions:**

**Fig. 1 | Boundary modes in a quantum Hall valley system. a,b**, Schematic of nematic domain walls between different broken valley symmetry quantum Hall states at effective filling factors $\tilde{v}=2$ (**a**) and $\tilde{v}=1$ (**b**), with Landau orbits denoted by the ellipses on either side. The expected valley flavor and degeneracy of the counter-propagating modes along the boundary depends on the filling factor as shown. **c**, Brillouin zone of the Bi(111), depicting multiple electron and hole valleys. **d**, Fermi surface in *k*-space of six-degenerate hole valleys, labeled to match different broken symmetry states discussed throughout the paper. Note the 90° rotation between the real-space Landau orbit orientation and *k*-space valley anisotropy. **e,f**, Gapless modes expected to connect topologically distinct phases on either side of the domain wall, with corresponding valley flavors and degeneracies indicated in the case of $\tilde{v}=2$ (**e**) and $\tilde{v}=1$ (**f**).

**Fig. 2 | Imaging a quantum Hall nematic domain wall. a, b**, Representative Landau level (LL) spectra far away from the domain walls, showing gaps arising from strain, $\Delta_{str}$, and exchange, $\Delta_{exch}$, which lift the six-fold valley symmetry. LL degeneracies labeled in parentheses. Changing the magnetic field allows us to tune the effective filling factor, $\tilde{v}$, from $\tilde{v}=2$ at $B = 14$ T, where exchange interactions split the four-fold degenerate multiplet around the Fermi level into two doubly degenerate states (**a**) to $\tilde{v}=1$ at $B = 13.4$ T, with one singly- and one triply-degenerate state around $E = 0$ (**b**). **c**, Differential conductance *dI/dV* map at the higher energy exchange split LL for $\tilde{v}=2$ ($E = 400$ μeV), showing a distinct boundary between two domains. Anisotropic Landau orbits pinned to atomic scale defects appear as low conductance elliptical rings that have a consistent orientation on either side of the domain wall. **d**, Large topography of pristine Bi(111) surface in the same field of view as (**c**),(**e**), with uniformity in height of better than 40 pm over the entire region. Dashed line denotes position of line cut in Fig. 3a,b. **e**, Conductance

map at the energy of the triply degenerate state for $\tilde{\nu}=1$ ($E = 330$ µeV), taken in the same area as (**c**),(**d**), which displays a domain wall of low conductance similar to (**c**). All measurements performed with setpoint voltage $V_{bias} = -400$ mV and setpoint current $I_{set} = 5$ nA.

**Fig. 3 | Spectroscopic characterization of topological boundary modes**. **a,b**, Spectroscopic line cuts along the line marked in Fig. 2d, showing the closing and re-opening of the exchange gaps as the topological invariant changes sign across the $\tilde{\nu}=2$ and $\tilde{\nu}=1$ domain walls, respectively. Individual spectra from the line cut at positions marked by colored triangles (red,blue) are shown in the corresponding side panels. **c,d**, Conductance maps at the Fermi energy between exchange-split LLs, showing enhanced local density of states at the domain wall, indicative of low energy boundary modes. **e,f**, Local density of states calculations for line cuts across a $\tilde{\nu}=2$ and a $\tilde{\nu}=1$ domain wall. **g**, The LL amplitude asymmetry in the vicinity of the $\tilde{\nu}=2$ domain wall arises from a dipole moment at the boundary, and qualitatively matches well to theoretical calculations. Black and green spectra are taken on the left and right sides of the domain wall, respectively, at locations marked by corresponding colored triangles in (**a**),(**e**).

**Fig. 4 | Interacting multi-channel boundary modes. a,b**, Individual spectra taken at the location of the domain wall. For $\tilde{\nu}=2$, a well-defined charge gap $\Delta_{charge} \approx 380$ µeV is observed (**a**), in contrast to a single peak at the $\tilde{\nu}=1$ domain wall (**b**). The behavior far from the domain wall is shown in the gray spectrum (reproduced from side panels in Fig. 3a and Fig. 3b). **c,d,e,f**, Possible interactions between the relevant valley-polarized edge modes account for the difference in insulating vs. metallic behavior at the domain wall. The four valley-flavored modes at the $\tilde{\nu}=2$ domain wall can be gapped by interactions of the form $A\bar{B} \rightarrow B\bar{A}$ (**c**), which reverses the chirality of the modes (backscatters them) in a manner where no net momentum is transferred in the surface Brillouin zone. Processes such as $AB \rightarrow \bar{A}\bar{B}$ (**d**) do not conserve 2D momentum

and are therefore exponentially suppressed. In the case of $\tilde{\nu}=1$ domain wall, which consists of two counter-propagating valley-polarized boundary modes, interactions are either gapless (AB → BA; (**e**)) or exponentially suppressed due to the large 2D momentum transfer between states (AA → BB; (**f**)).

**Methods**

**Sample Preparation and Measurement**

Single Bi crystals were grown using the Bridgman method from 99.999% pure Bi that had been treated to remove oxygen impurities. The samples were cleaved in ultrahigh vacuum at room temperature, immediately inserted into a home-built dilution refrigerator STM and cooled to cryogenic temperatures. Except where noted, all measurements were performed at 250 mK using a W tip. Spectra and conductance maps were acquired using a lock-in amplifier with AC rms excitation $V_{rms}$ = 30 µV or $V_{rms}$ = 74 µV. The setpoint bias voltage was $V_{set}$ = -400 mV and the setpoint current was $I_{set}$ = 5 nA.

**Landau Levels of the Bi(111) Surface**

In the presence of a large magnetic field $B$, the Bi(111) surface states (see Fermi surface in Fig. 1c) are quantized into Landau levels (LLs), resulting in sharp peaks in the STM spectra. We use the evolution of these states as a function of magnetic field to distinguish between electron- and hole-like LLs, which disperse in energy with a positive or negative slope, respectively. We find that the surface charge density is constant, allowing us to tune the filling factor of a Landau level by changing magnetic field[16]. The six-fold valley degeneracy of the hole-like states can be lifted both by strain and exchange interactions. Strain is a single particle effect that opens up a gap, which varies by location but is independent of magnetic field and affects the hole-like LLs regardless of their energy. In contrast, exchange interactions result in spontaneous symmetry breaking that lifts the LL degeneracy only around the Fermi level ($E$=0)[16,17]. In this location of the sample, the strain-induced splitting, gives rise to a two- and four-fold degenerate multiplet for the LL with orbital index $N$ = 3, marked as $\Delta_{str}$ in Fig. 2a,b, and Extended Data Fig. 1a. As the four-fold degenerate multiplet is tuned to the Fermi level by varying magnetic field, an additional exchange gap $\Delta_{exch}$ opens, as shown in Fig. 2a,b (taken in the same area as Extended Data Fig. 1a). The difference in relative amplitude of the two exchange-split LLs as a function of magnetic field (compare Fig. 2a to 2b), as well as the conductance maps (Fig. 2c,d) are used to identify that a single valley state is occupied at $B = 13.4$ T (effective filling factor $\tilde{v}=1$) whereas a pair of valleys with the same anisotropy are occupied at $B = 14$ T ($\tilde{v}=2$).

Spectroscopic mapping of variations in the differential conductance can be used to image individual Landau orbits, seen as concentric ellipses of suppressed conductance pinned to defects on the sample surface. The sharp potentials of atomic scale defects shift one of the cyclotron orbits (with guiding center index $m = N$) up or down in energy depending on the sign of the defect potential. Conductance maps at the LL energies display this missing orbit, where the number of concentric ellipses reflect the number of antinodes of the wavefunction. The ellipticity of the Landau orbits results from the effective mass anisotropy of the valleys, with the usual 90° rotation in orientations between real space and momentum space.

**Role of Interactions in Domain Wall Formation**

Interactions play a central role in the formation of the domain wall, corroborated by the absence of two inequivalent nematic regions in measurements under conditions where exchange effects are absent. At a magnetic field corresponding to $\tilde{v}=0$, where the four-fold degenerate LL completely unoccupied (i.e. tuned away from the Fermi level) and is not split by exchange (Extended Data Fig. 1a), a conductance map taken in the same field of view as the maps in Fig. 2 shows Landau orbits of both orientations throughout the entire image (Extended Data Fig. 1b). In this regime, the symmetry between these two orientations is not broken and no domain wall is present. The absence of domains is further highlighted by the line cut of spectra in Extended

Data Fig. 1c, which crosses the original position of the domain wall, but shows that the LL remains four-fold degenerate throughout. These observations confirm that the domains form as a result of exchange interactions, although the strain likely provides a bias that determines which valley(s) the QHFM spontaneously occupies.

Furthermore, the exchange interactions are suppressed and the LL remains four-fold degenerate when the temperature is raised from 250 mK to 2K, for the same magnetic field which produced the $\tilde{\nu}=2$ domain wall (spectra in Extended Data Fig. 1d). A conductance map at this higher temperature, taken in the same region as Fig. 2, shows two superimposed orientations of Landau orbits (inset Extended Data Fig. 1e), in stark contrast to the nematic domains at base temperature. A corresponding line cut (Extended Data Fig. 1f) across the initial domain wall location demonstrates that we cannot resolve exchange splitting between the two valley orientations. It is possible that the absence of splitting is a sign that the sample is above the critical temperature of the nematic transition or it could reflect a decreased energy resolution from thermal broadening. Regardless, these data show that the domain wall does not exist in the absence of electronic interactions.

**Role of Strain in Domain Wall Formation**

Although we show above that this nematic domain wall forms only in the presence Coulomb interactions, it is likely that a local strain field stabilizes the position of the domain wall. Strain is a natural candidate that couples to the valley degree of freedom. Extended Data Fig. 2a shows a schematic of a possible strain field that gives rise to the particular valley splittings we observe. Specifically, there is a large strain in the direction of valleys $C$ and $\bar{C}$, which remains relatively unchanged across the domain wall and lowers the energy of these two valleys compared to the other four. However, in the presence of exchange interactions, a small switch in orientation of the strain field from favoring valleys A and $\bar{A}$ to favoring valleys B and $\bar{B}$ provides a symmetry-breaking perturbation which gives rise to a nematic domain wall. An experimental linecut across the $\tilde{\nu}$=2 domain wall (Extended Data Fig. 2b) shows that the energy of the two-fold degenerate valley state (at $E$ ~ -1.25 meV corresponding to valleys C and $\bar{C}$), which is split off from the other valley states by strain, does not change significantly across the domain wall associated with valleys A, $\bar{A}$ and B, $\bar{B}$.

**Properties of Domain Wall Boundary Modes**

The quantum Hall valley states are intrinsically topological and therefore any crossing of valley states (either due to exchange or strain) would create domain walls between different quantum Hall states with boundary modes that are topological. Theoretically, we expect the boundary between two topologically distinct domains that are valley-polarized to host counter-propagating modes. Experimentally, we directly demonstrate the presence of low energy modes at such domain walls that form spontaneously and visualize the difference in nematic order on either side. However, our experiments thus far do not probe the counter-propagating or the valley-polarization properties of the boundary modes themselves. Nevertheless, our spectroscopic measurements of this interacting 1D system, which shows drastically different behavior at the domain wall depending on the valley flavor of the boundary modes (insulating for $\tilde{\nu}$=2 and metallic for $\tilde{\nu}$=1), supports the existence of counter-propagating valley-polarized modes. We note that if the LL crossing occurs away from the Fermi level, interactions cannot affect the properties of the boundary modes as all the states are either occupied or unoccupied. We emphasize that the role of interactions is not to render boundary modes trivial or topological (they are all topological when they occur in such a quantum Hall system) but rather to change the properties of these modes.

We note the slight differences in the meanderings of the two domain walls, where the $\tilde{\nu}=2$ boundary is straighter that that of $\tilde{\nu}=1$. It is possible that the charge gap for $\tilde{\nu}=2$ makes it more rigid, whereas the metallic nature of the $\tilde{\nu}=1$ domain wall makes it more accommodating to distortions. However, addressing the cause of this difference in shape would require further investigation into the stiffness of each of these domain walls.

**Differential Conductance Maps at Additional Energies**

The experimental measurements of differential conductance *dI/dV* presented in Fig. 2-3 demonstrate the electronic behavior of a nematic domain wall for two key energies: at the higher energy exchange-split LL and at the Fermi level for both the $\tilde{\nu}=1$ and $\tilde{\nu}=2$ domain walls. These data clearly show the presence of two distinct regions with different broken rotational symmetry as well as low-energy states at the boundary between them. To further illustrate the evolution of electronic properties of these domain walls, we show additional conductance maps taken in the same area of the sample.

Extended Data Fig. 3 displays maps for the $\tilde{\nu}=2$ domain wall, taken at $B = 14$ T, with approximately 100 μeV energy spacing. Comparing Extended Data Fig. 3a to Extended Data Fig. 3i highlights the switch in the Landau level wavefunction orientation and the corresponding valleys between the occupied and unoccupied exchange split states. These conductance maps in conjunction with the exchange splitting of a four-fold degenerate LL into two doubly degenerate states as shown in Fig. 2a establish that the domains occur from pairs of valleys at opposite momenta, specifically valleys A and $\bar{A}$ in the left region and valleys B and $\bar{B}$ to the right of the domain wall in Extended Data Fig. 3i. Furthermore, as the energy is decreased from 400 μeV, the inward evolution of the regions of high differential conductance is a manifestation of the exchange gap closing in the vicinity of the domain wall.

We note here that in Extended Data Fig. 3d,e, the orientation of the high conductance Landau orbits (i.e. those shifted in energy from their respective LLs due to defect potentials) is not fully uniform within each domain. This reflects the fact that certain defects shift the Landau orbit to higher energy, while others shift it to lower energy. As a result, for measurements at energies between the LL peaks, cyclotron orbits shifted upward in energy from the lower peak and downward in energy from the upper peak are both simultaneously visible. We emphasize that the presence of both directionalities at these energies does not reflect any imperfection in the domains, whose uniformity is clear from Extended Data Fig. 3a and Extended Data Fig. 3i.

Changing the magnetic field to $B = 13.4$ T allows us to tune the filling factor of the LLs to $\tilde{\nu}=1$, where the four degenerate LL is split into one singly- and one triply-degenerate state (Fig. 2b). Conductance maps under these conditions at several different energies are shown in Extended Data Fig. 4. The three-fold degenerate LL corresponds to three of the valleys, and accordingly, conductance maps at approximately that energy (Extended Data Fig. 4c,d), show dark Landau orbits of two orientations on either side of the stripe of low conductance that marks the domain wall. In either domain, one of the two orientations is more pronounced, in agreement with a triply degenerate LL, which arises from two valleys with the same anisotropy and a third valley with a different anisotropy direction. The favored directionality switches across the boundary as consistent with a nematic domain wall (see also Fig. 2e). Moreover, conductance maps measured at the energy of the singly degenerate LL show wavefunctions of only one orientation within each region (Extended Data Fig. 4a), whose directionality matches the weaker direction in Extended Data Fig. 4c,d and changes across the boundary. At this energy, there is also residual spectral weight from the tails of the gapless topological boundary mode (refer to Fig. 3b and Fig. 4b for the width of the peak), which results in the enhanced boundary conductance that tracks the domain wall at $E = $ -120 µeV.

The map at the Fermi level in Extended Data Fig. 4b reveals increased conductance between domains, indicating the presence of low-energy edge modes at the domain wall. In the case of $\tilde{v}=1$, a single hole valley switches its occupation with another one that has a different anisotropy orientation. Our previous work established that a singly degenerate LL in this system is valley polarized, however, we cannot experimentally distinguish which specific valley is occupied[17]. Without loss of generality, we label the two states that cross at the $\tilde{v}=1$ domain wall by valleys A and B.

**Robust Metallic vs. Insulating Behavior at Domain Wall**

We present additional measurements of linecuts across the $\tilde{v}=2$ domain wall (Extended Data Fig. 5a-f) and across the $\tilde{v}=1$ domain wall (Extended Data Fig. 5g-l). The variation in the spectra are likely due to disorder in the samples.

Extended Data Fig. 6 demonstrates that the difference in electronic behavior at the two filling factors is a robust feature along the domain walls in each case. Individual spectra from several locations at the domain wall exhibit an interaction-driven charge gap at the $\tilde{v}=2$ domain wall (Extended Data Fig. 6a), in contrast to the gapless spectra in the case of a $\tilde{v}=1$ domain wall (Extended Data Fig. 6b). The spectra at the $\tilde{v}=2$ domain wall exhibit a charge gap, which ranges from 325 µeV to 425 µeV, as also seen in a linecut along the boundary (Extended Data Fig. 7). The greater variation in spectra for $\tilde{v} = 2$ at the domain wall is possibly due to a gap enhancement from atomic defect backscattering, which can further localize valley-polarized edge modes through inter-valley scattering. However, the isolated effect of individual defects is minimal and does not open a resolvable gap in spectra at the $\tilde{\nu}=1$ domain wall, which is measured with the same disorder potential as the $\tilde{\nu}=2$ domain wall, thus supporting our claim that this charge gap occurs due to Coulomb interactions constrained by valley flavor.

**Data Availability**

The data that supports the findings of this study are available from the corresponding author upon reasonable request.

**Extended Data Figure Captions:**

**Extended Data Fig. 1 | Role of electron-electron interactions in domain wall formation. a**, Spectrum away from the domain wall at $B = 13.1$ T and $T = 250$ mK, where the effective filling factor $\tilde{\nu} = 0$, so the four-fold degenerate LL is not split by exchange. **b**, Conductance map at the four-fold degenerate LL peak energy $E = 700$ μeV in the same area as in Fig. 2. The presence of cyclotron orbits of both orientations throughout the image indicates the absence of a domain wall under these conditions. **c**, Spectroscopic line cut along the dashed line in **(b)** also show a four-fold degenerate LL that does not change across the original location of the domain wall, in stark contrast to the line cut in Fig. 3a. **d**, Spectrum away from the domain wall at $B = 14$ T corresponding to the $\tilde{\nu}=2$ domain wall but at a higher temperature of $T = 2$ K. Again, we do not resolve any exchange splitting and the LL is four-fold degenerate. **e**, Topography of the same area (identical to Fig. 2d) overlaid with a $dI/dV$ map at $E = -100$ μeV. We observe cyclotron orbits of both orientations at this elevated temperature, and no domain wall is visible. **f**, Line cut

along the dashed line in (**e**). The absence of splitting in the four-fold degenerate LL confirms that the domain wall is not present at 2 K.

**Extended Data Fig. 2 | Schematic of strain field and comparison to experimental linecut. a**, Schematic of possible strain field (top) and the resulting energies of the different valley states (bottom). A large component of the strain in the direction of valleys $C$ and $\bar{C}$ lowers the energy of these two valleys compared to the other four. In the presence of exchange interactions, the switch in strain field from a slight favoring of valleys $A$ and $\bar{A}$ to valleys $B$ and $\bar{B}$ gives rise to the nematic domain wall. **b**, Experimental linecut across the $\tilde{\nu}=2$ domain wall showing that the energy of the two-fold degenerate valley state (corresponding to $C$ and $\bar{C}$ at $E \sim -1.25$ meV is split off from the other valley states by strain, and does not change significantly across the domain wall associated with the crossing between pairs of valleys $(A, \bar{A})$ and $(B, \bar{B})$ at the Fermi level.

**Extended Data Fig. 3 | Energy-dependence of the domain wall behavior at $\tilde{\nu} = 2$. a-i**, Differential conductance maps measured in the same location and under identical conditions to those in Fig. 2c. Each panel shows *dI/dV* at a different energy, ranging from the lower-energy exchange-split LL at $E = -400$ μeV (**a**) to the higher-energy LL peak at $E = 400$ μeV (**i**). The data demonstrate the different preferred wavefunction orientations for each respective domain as well as the different orientations of occupied and unoccupied states within a given domain.

**Extended Data Fig. 4 | Energy-dependence of the domain wall behavior at $\tilde{\nu} = 1$. a-d**, Differential conductance maps measured in the same location and under identical conditions to those in Fig. 2e. Each panel shows *dI/dV* at a different energy, ranging from the singly degenerate LL at $E = -120$ μeV (**a**) to the triply degenerate LL peak at $E = 330$ μeV (**d**).

**Extended Data Fig. 5 | Additional spectroscopic line cuts across the domain wall. a-f**, Spectroscopic line cuts across the $\tilde{\nu} = 2$ domain wall. The six different line cut trajectories are indicated by the dashed lines in (**m**). While minor variations in the spectra are seen, likely due to the effects of local disorder, the key features of exchange gap closing and LL crossing at the domain wall are consistent. **g-l**, Spectroscopic line cuts across the $\tilde{\nu} = 1$ domain wall. Again, the same features of the change in topological invariant are present in each line cut. **m,n**, *dI/dV* maps reproduced from Fig. 2c and Fig. 2e, respectively, overlaid with dashed lines showing the locations of the spectroscopic line cuts in (**a**)-(**l**).

**Extended Data Fig. 6 | Variation in individual spectra at domain wall. a**, Individual spectra (blue, green, brown, purple, black, red) measured for $\tilde{\nu} = 2$ (**a**) and $\tilde{\nu} = 1$ (**b**), at domain wall positions corresponding to Line 1 to Line 6 (Extended Data Fig. 5). All spectra in (**a**) show a charge gap $\Delta_{charge}$ that is smaller than the exchange gap far from the domain wall (grey dashed spectrum). No LL splitting is visible at the $\tilde{\nu} = 1$ domain wall (**b**), in contrast to the behavior in (**a**) and far from the domain wall (grey dashed spectrum).

**Extended Data Fig. 7 | Linecut along $\tilde{\nu} = 2$ domain wall. a**, Linecut parallel to the $\tilde{\nu} = 2$ domain wall, along the white line in (**b**), showing spatial variation in the spectra **b**, Conductance map reproduced from Fig. 2c with the location of the spectroscopic linecut in (**a**) marked by the white line.

Figure 1

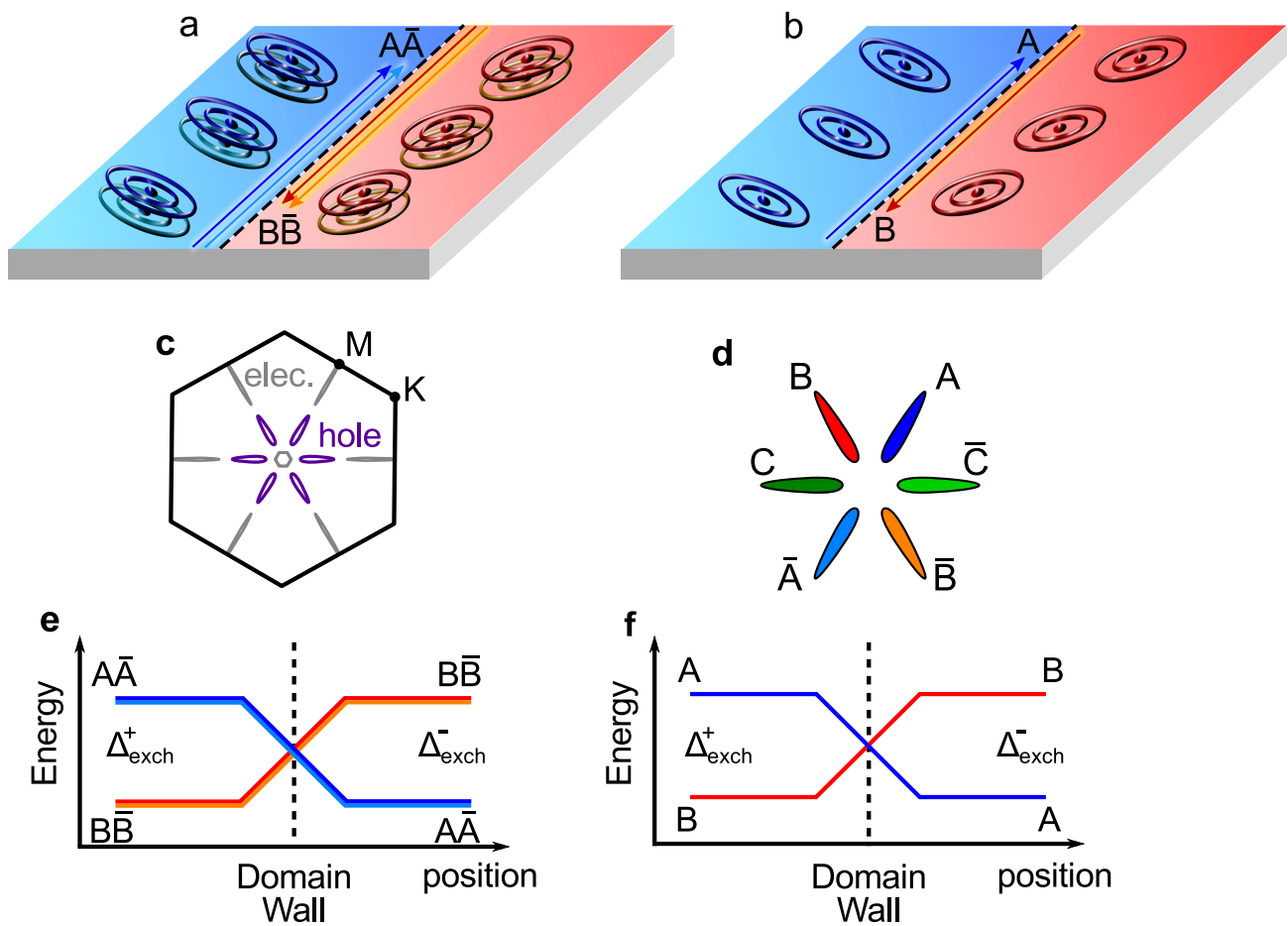

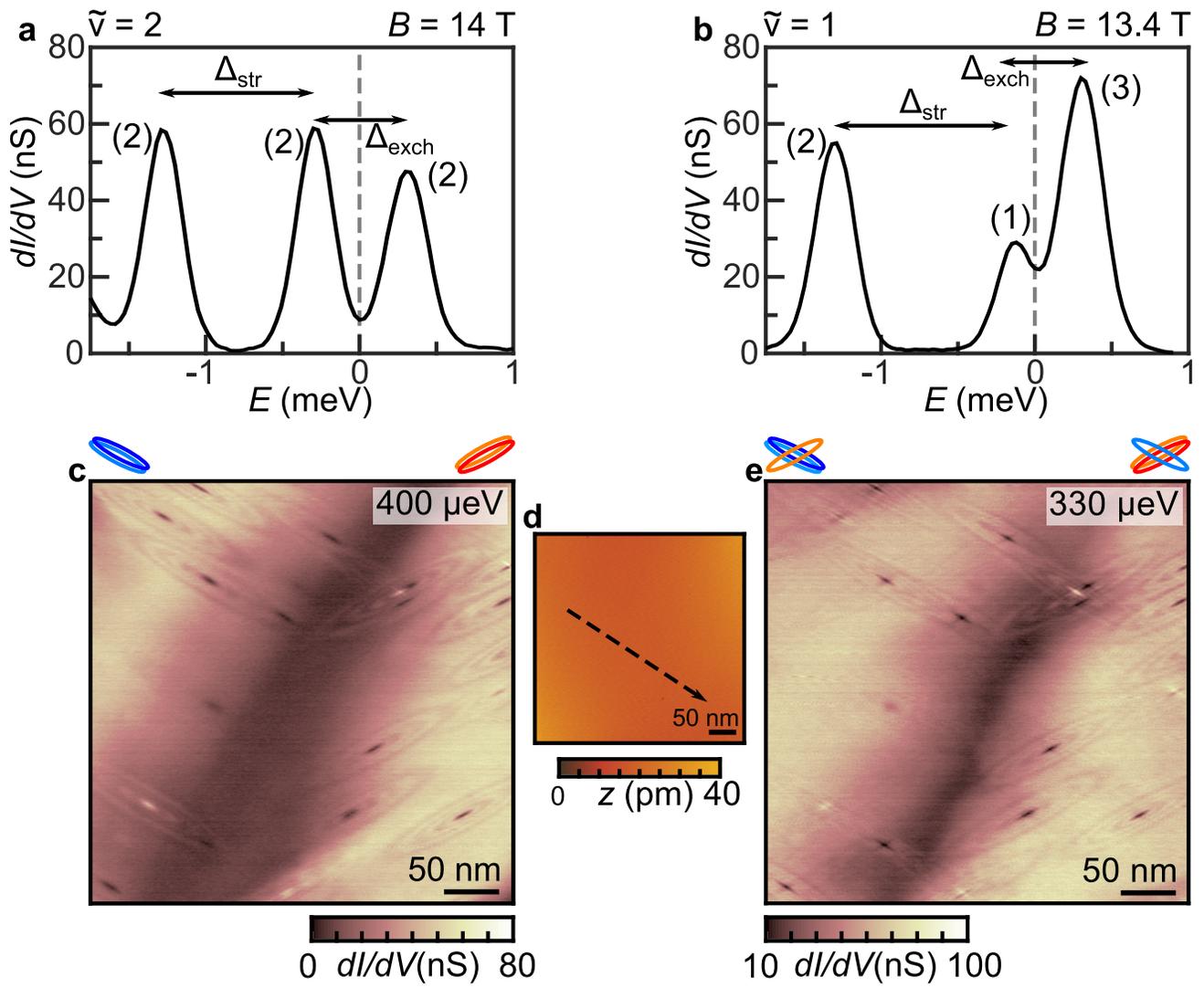

# Figure 3

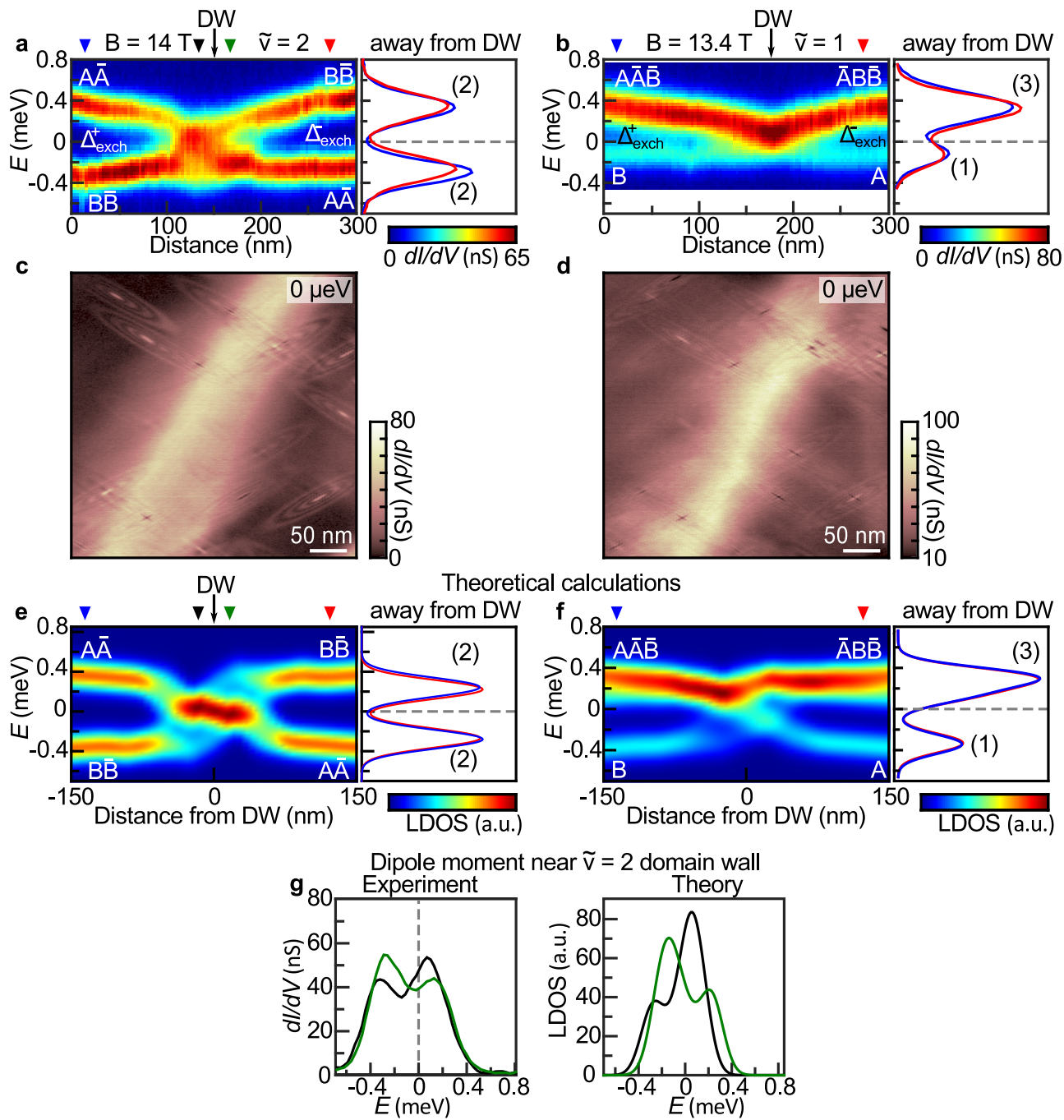

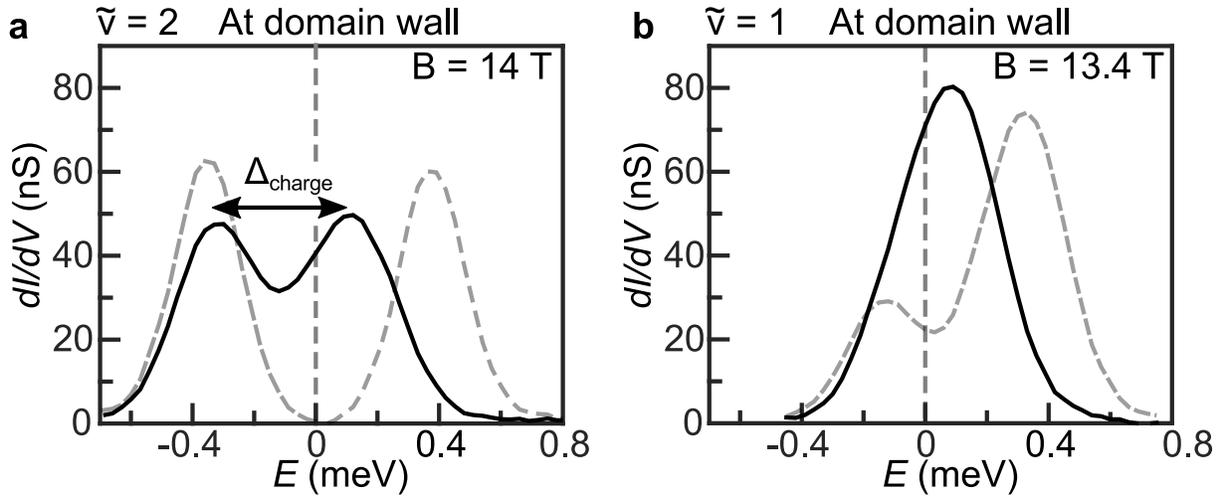

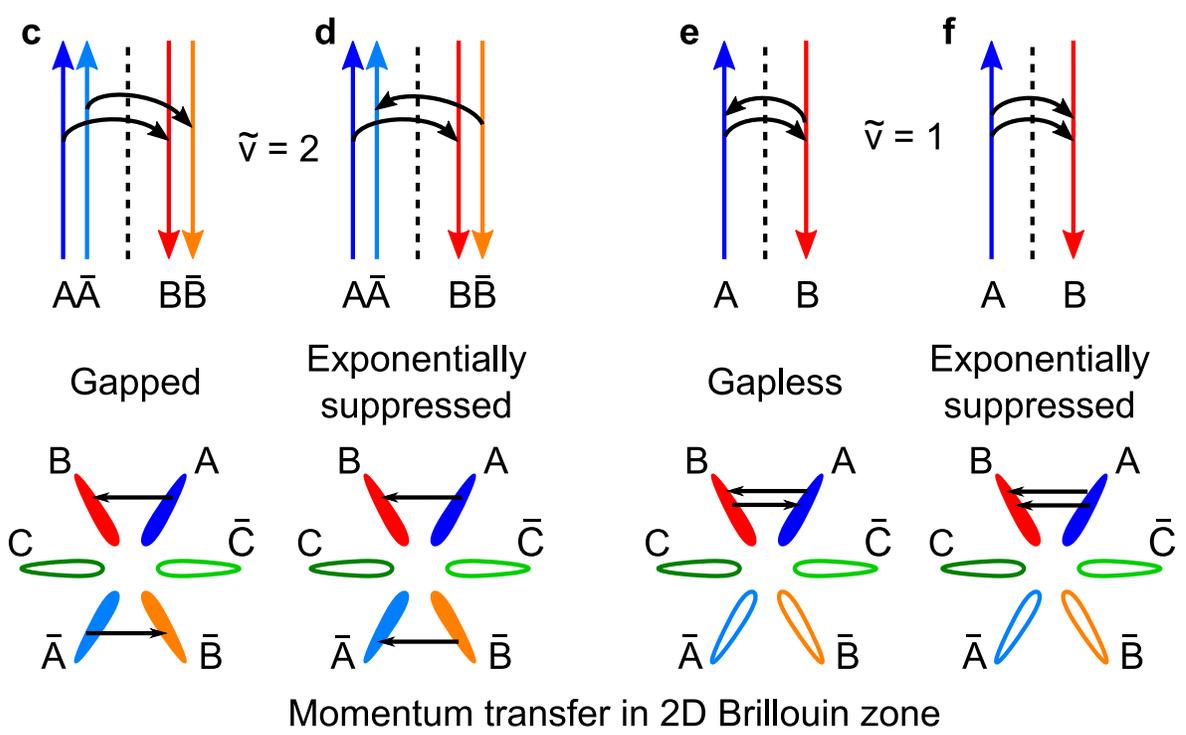

Momentum transfer in 2D Brillouin zone

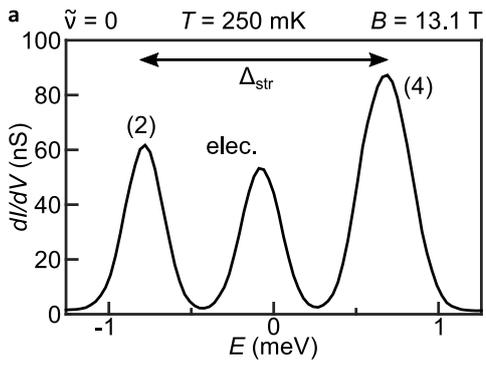
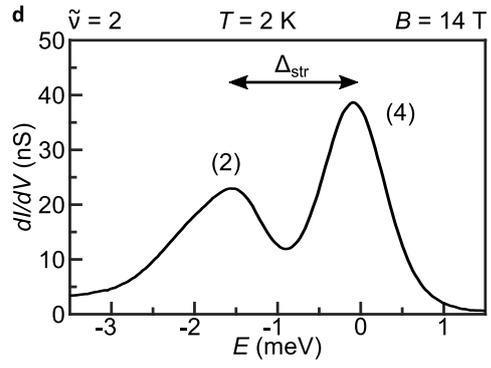
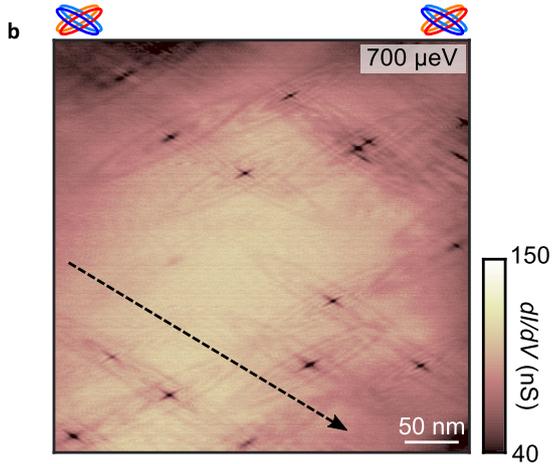
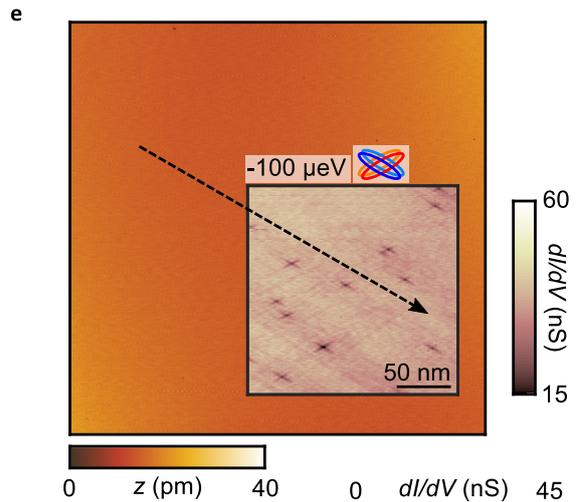
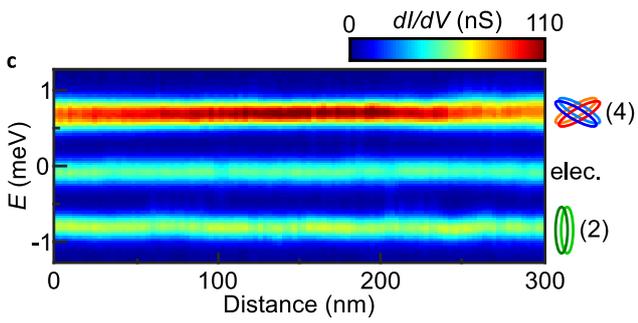
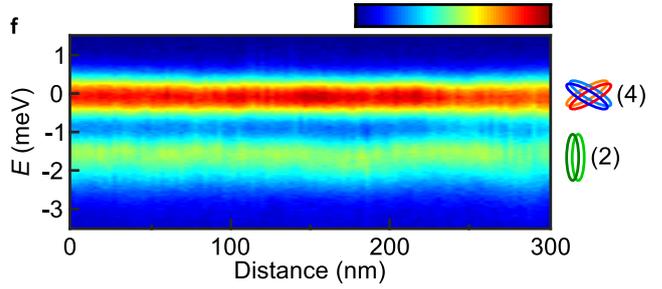

# Extended Data Figure 2

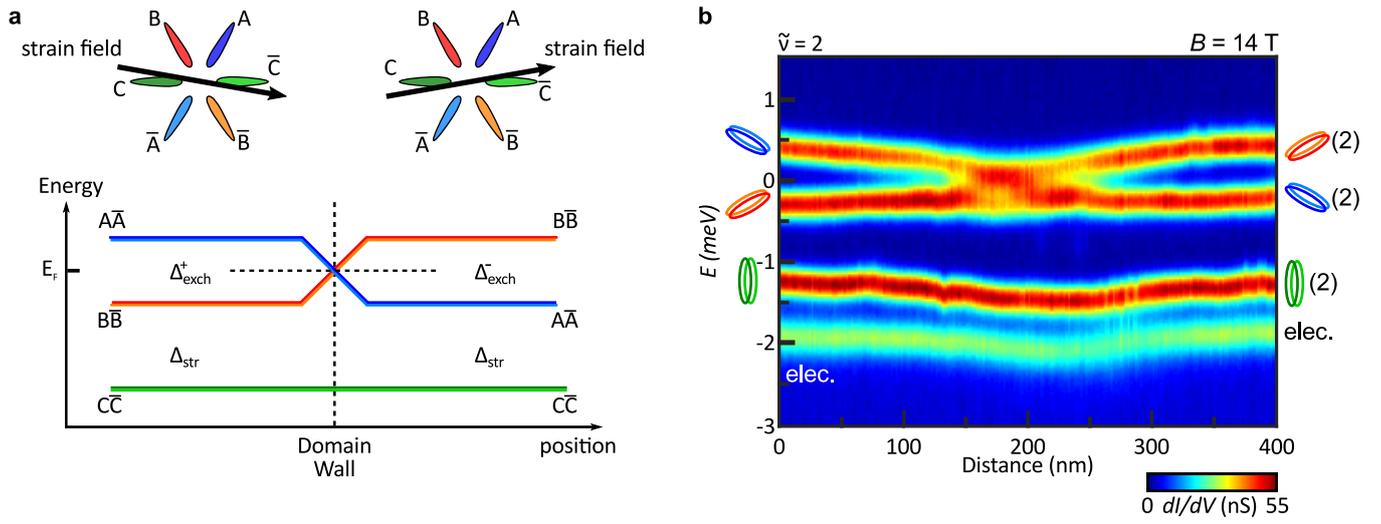

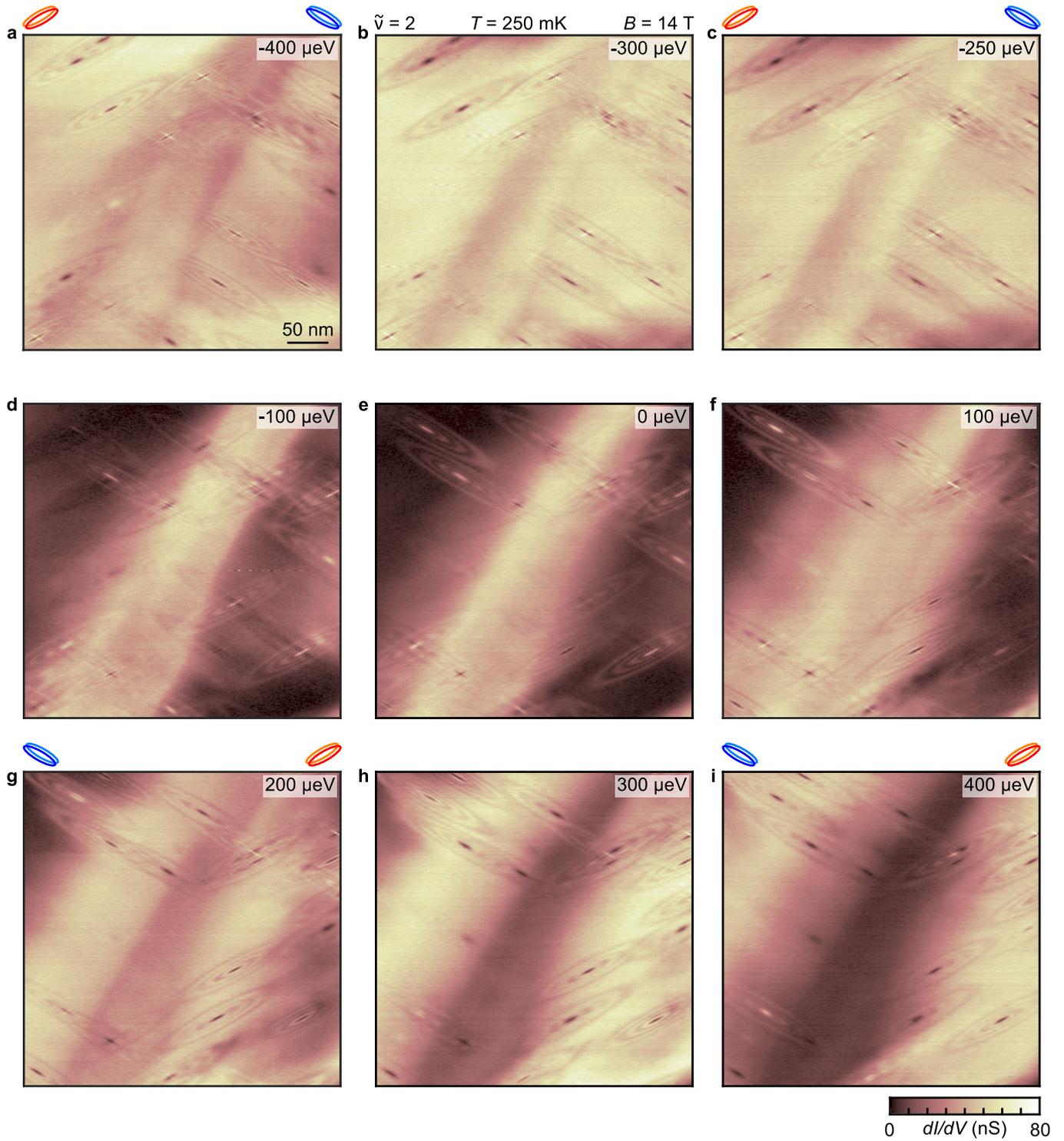

# Extended Data Figure 4

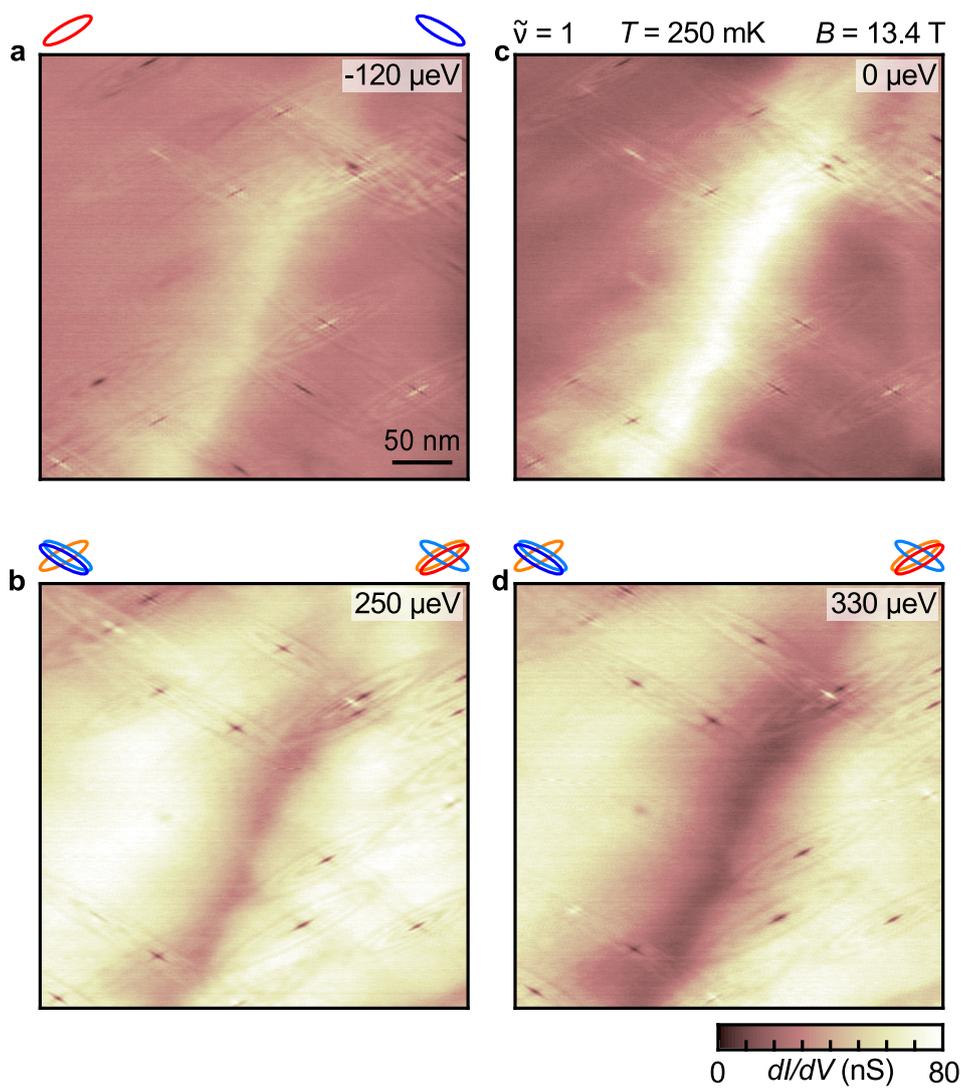

# Extended Data Figure 5

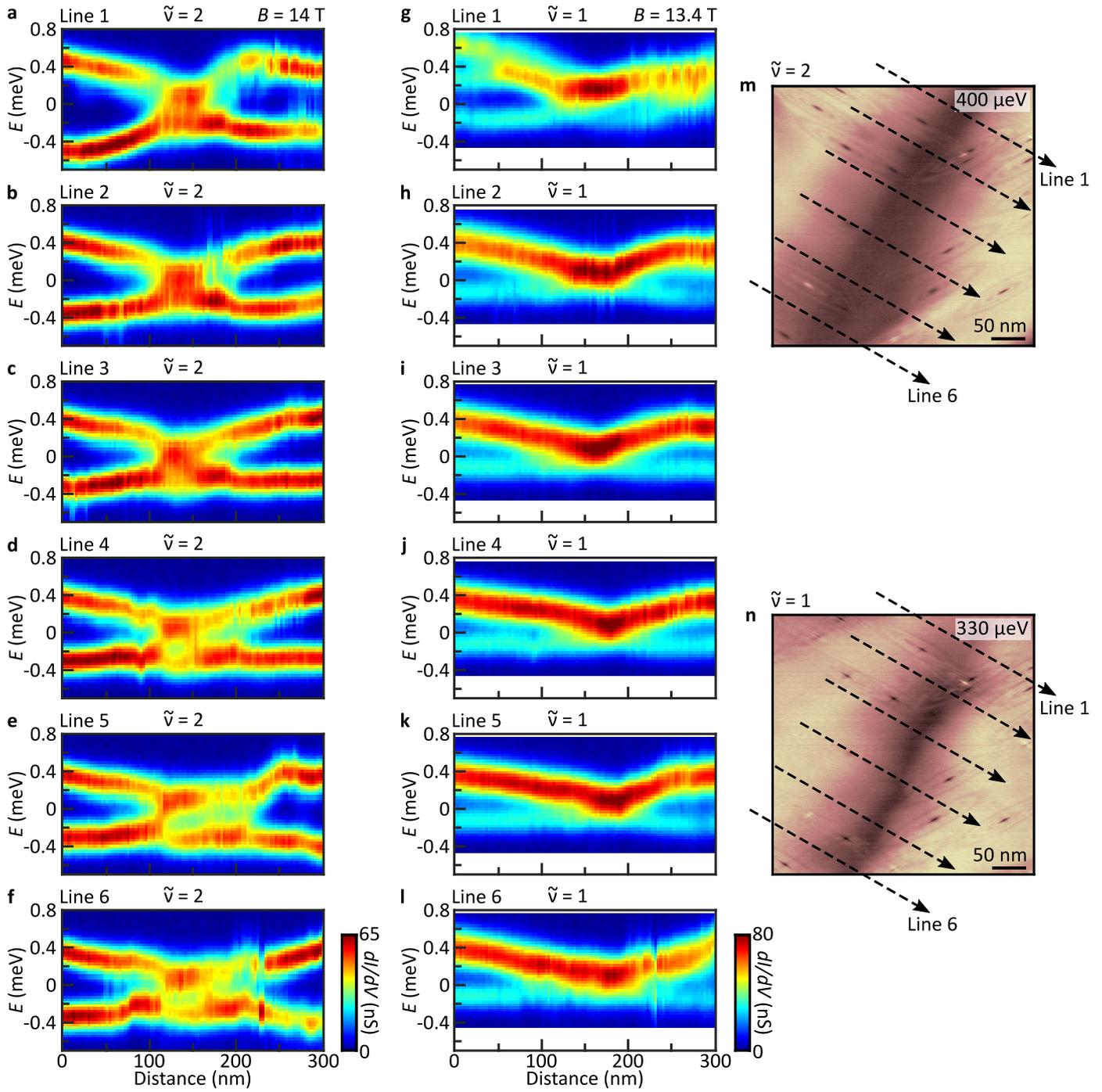

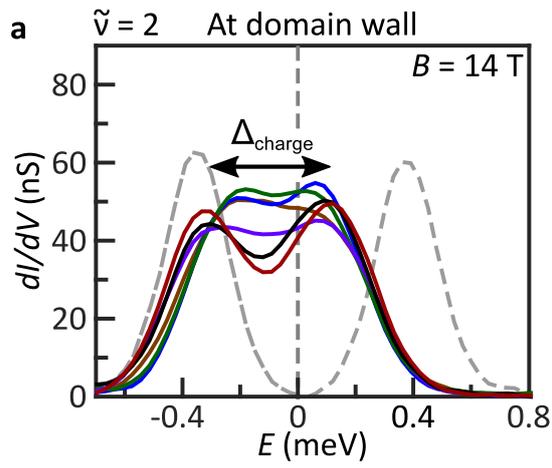
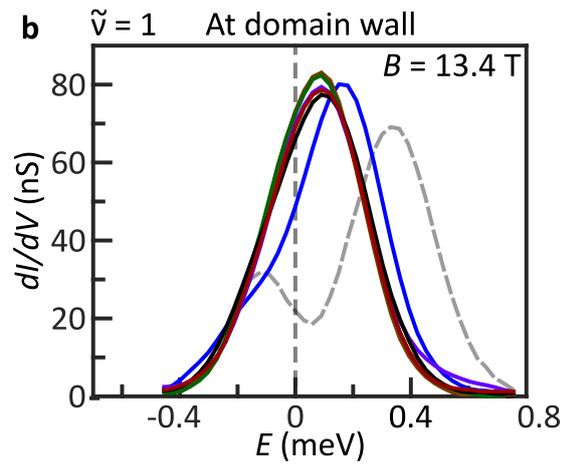

## Extended Data Figure 7

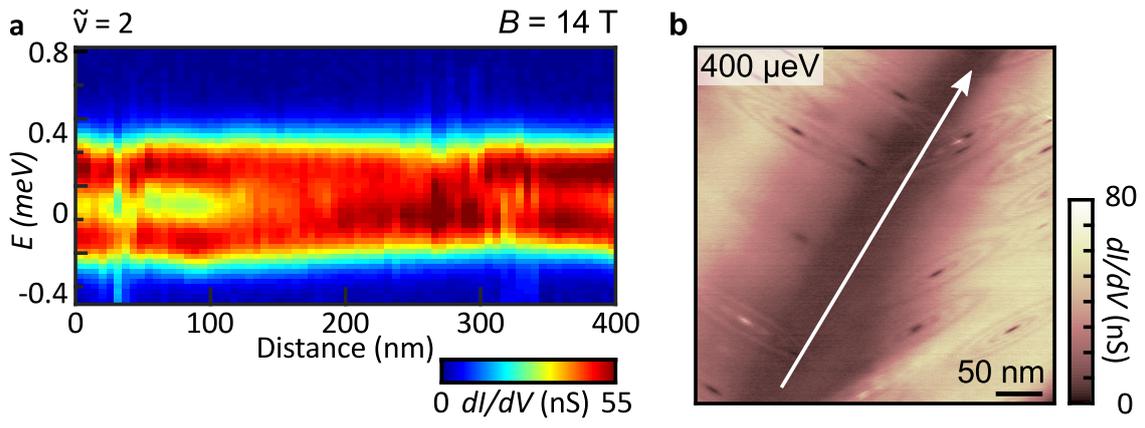